# The Role of Computational Modeling in Enhancing Thermal Safety During Cardiac Ablation

**Leila Seidabadi[1], Indra Vandenbussche[1], Rowan Carter Fink, MacKenzie Moore, Bailey McCorkendale, Fateme Esmailie[*]**

Department of Biomedical Engineering, University of North Texas, Texas, USA

* Corresponding author. University of North Texas Discovery Park, 3940 North Elm Street, Denton, Texas 76207-7102, USA.

Tel: +1(940)3698988; E-mail: Fateme.Esmailie@unt.edu

[1] Co-first authors

Word Count: 6365

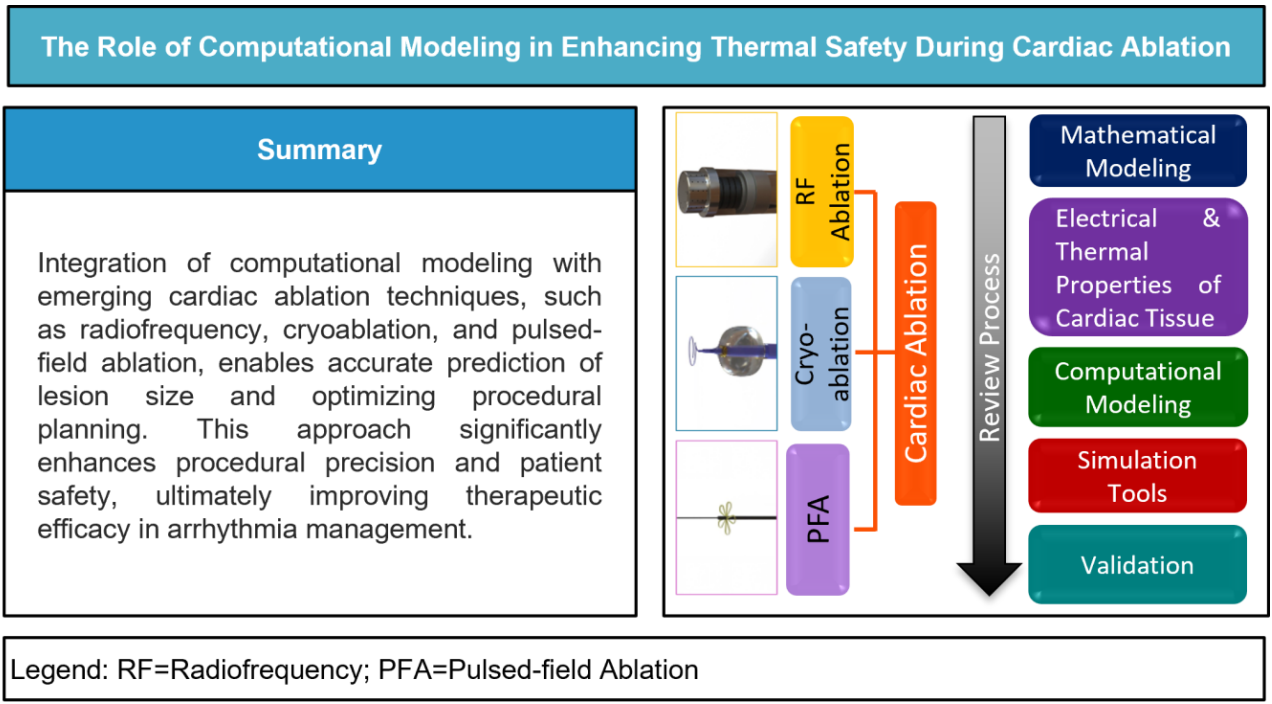

The Role of Computational Modeling in Enhancing Thermal Safety During Cardiac Ablation

**Summary**

Integration of computational modeling with emerging cardiac ablation techniques, such as radiofrequency, cryoablation, and pulsed-field ablation, enables accurate prediction of lesion size and optimizing procedural planning. This approach significantly enhances procedural precision and patient safety, ultimately improving therapeutic efficacy in arrhythmia management.

Legend: RF=Radiofrequency; PFA=Pulsed-field Ablation

## Abstract

**Objective:** In this review, we aim to provide an analysis of current cardiac ablation techniques, such as radiofrequency ablation (RF), cryoablation, and pulsed-field ablation (PFA), with a focus on the role of computational modeling in enhancing the precision, safety, and effectiveness of these treatments. Particular attention is given to thermal management, exploring how computational approaches contribute to understanding and controlling energy delivery, heat distribution, and tissue response during ablation procedures.

**Methods:** The mechanisms, applications, and limitations of radiofrequency (RF) ablation, cryoablation, and pulsed field ablation (PFA) are reviewed. Additionally, the use of computational approaches, including numerical methods and artificial intelligence (AI)-based models, for evaluating energy distribution, lesion size, and tissue response during ablation procedures is discussed.

**Results:** Computational methods can predict ablation treatment outcomes and help optimize lesion size, ablation parameters, and procedural safety. However, these models are only reliable when properly validated and verified.

**Conclusion:** Further research is essential to collect reliable *in vivo* data for validating computational models and integrating them into clinical practice to improve patient outcomes.

| | |
|---|---|
| **ABBREVIATIONS** | |
| AF | Atrial Fibrillation |
| AI | Artificial Intelligence |
| CE | Contrast-Enhanced |
| CFD | Computational Fluid Dynamic |
| CT | Computed Tomography |
| DE | Delayed-Enhanced |
| DF | Dominant Frequency |
| ECG | Electrocardiogram |
| E-PVI | Empirical Pulmonary Vein Isolation |
| FDM | Finite Difference Method |
| FEM | Finite Element Method |
| FVM | Finite Volume Method |
| LBM | Lattice Boltzmann Method |
| LGE | Late Gadolinium Enhance |
| ML | Machine Learning |
| MRI | Magnetic Resonance Imaging |
| PFA | Pulsed-Field Ablation |
| PVI | Pulmonary Vein Isolation |
| RF | Radiofrequency |
| TID | Thermal Iso-effective Does |
| V-DF | Virtual Dominant Frequency |

## 1. Introduction

The heart's conduction system is essential for effective cardiovascular function. Disruptions in the conduction system can lead to arrhythmias, which are characterized by irregular heartbeats that may manifest as tachycardia, bradycardia, or other dysrhythmias [1]. These disturbances pose significant health risks and may require treatment in the form of medication, lifestyle changes, cardioversion, or surgical interventions with cardiac ablation emerging as a key method for restoring normal sinus rhythm [2].

Cardiac ablation techniques can be categorized into several modalities, each with distinct mechanisms and applications for treating arrhythmias. Radiofrequency Ablation (RF) [3] and

Cryoablation [4] are the most widely used methods, which induce thermal damage to disrupt the arrhythmic conduction pathway [5]. RF utilizes radiofrequency waves to heat the tissue (60°C) [6], and cryoablation employs gases to achieve extremely low temperatures (below -40°C), with the gases remaining confined within the catheter and not being delivered to the tissue [7]. Pulsed Field Ablation (PFA) [8] represents a novel approach that uses high voltage pulsed electric fields to induce Irreversible Electroporation (IRE) and does not rely on thermal energy.

Alcohol ablation is used for cardiomyopathy by injecting ethanol into heart arteries; however, it carries higher hospital complication rates and uncertain long-term survival outcomes [9]. Other techniques, such as, laser ablation [10], ultrasound ablation [11], microwave ablation [12], and thermal balloon ablation [13], are rarely applied in cardiac settings due to their complexity and the expertise required (Fig. 1) [14]. Thus, these methods are not discussed in this review paper. Cryoablation, RF, and PFA, as the primary methods used in cardiac ablation, are discussed in detail (Fig. 1).

In addition to advancements in ablation techniques, the use of computational modeling to design ablation strategies and predict procedural outcomes has increased, aiming to enhance the precision and safety of cardiac ablation procedures [14]. Mathematical models and simulation tools can be used to predict temperature and electric field distribution within tissues and evaluate the extent of lesion formation, thus, enabling patient-specific treatment planning and real-time decision-making [15].

In this paper, we present an overview of current cardiac ablation techniques and their associated computational models for evaluating thermal effects.

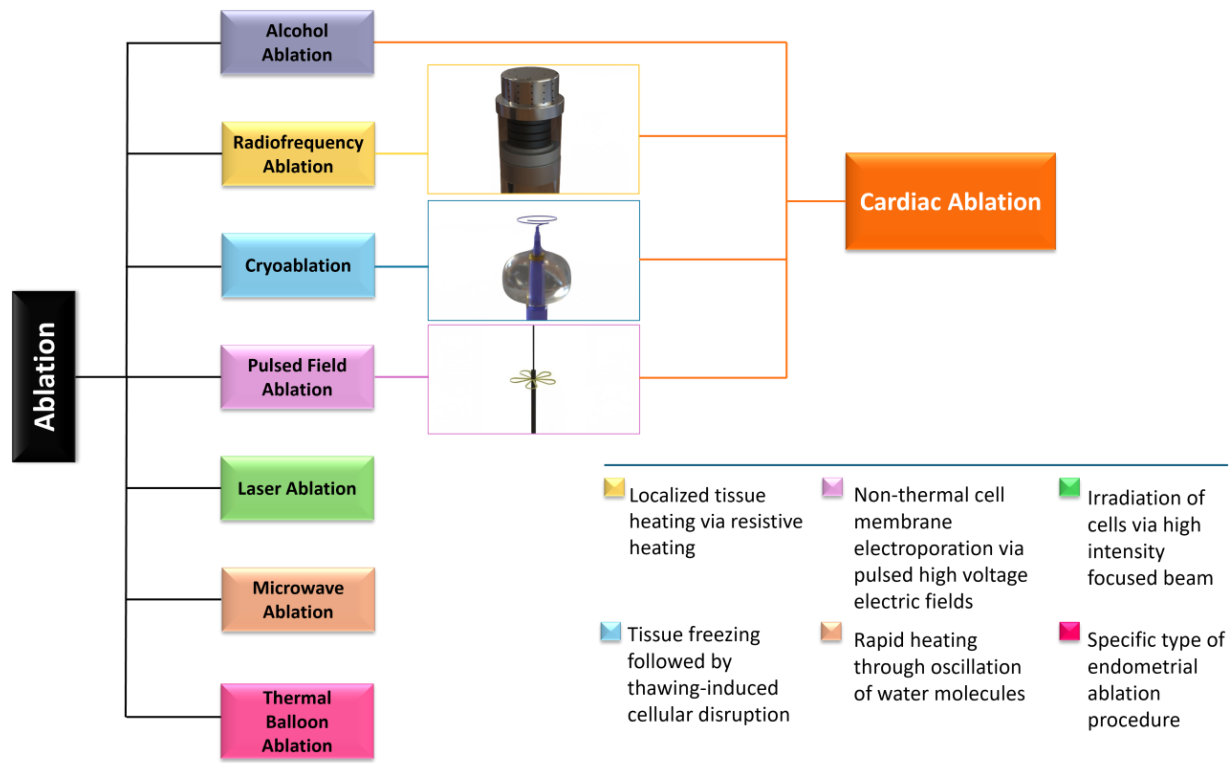

Figure 1: Schematic of ablation techniques

## 2. Overview of cardiac ablation techniques

Arrhythmias can arise from a range of physiological factors, including genetic predispositions, structural heart changes, electrolyte imbalances, and the influence of medications or stimulants [2]. In addition, pathological conditions such as ischemic heart disease, cardiomyopathy, and heart valve disorders can contribute to the development of arrhythmias [16].

Ablation techniques, particularly catheter ablation, target the areas of the heart responsible for initiating or sustaining abnormal electrical activity [17]. By precisely creating lesions to disrupt faulty electrical pathways, ablation procedures help restore normal heart rhythm, alleviate symptoms, and enhance overall cardiac function [18].

Multiple ablation techniques have been used to address different types of arrhythmias [19]. Among various cardiac ablation methods, RF ablation remains the predominant choice for cardiac ablation, while cryoablation and PFA are utilized in more specific cases or as emerging alternatives [3]. Using RF ablation offers significant advantages, including high precision in lesion formation and proven effectiveness in maintaining sinus rhythm, particularly in atrial fibrillation patients [5]. The introduction of multielectrode irrigated catheters has further refined its application, enhanced the accuracy of lesion placement and minimized damage to surrounding tissues [20–22]. Despite RF ablation's well-recognized therapeutic benefits, it has faced several limitations, such as: 1- the risk of post-ablation thromboembolism, 2- difficulties in accessing deeper tissues, 3- unintended damage to adjacent vascular and electrical structures, and 4- challenges in evaluating electrophysiologic effects before causing irreversible local tissue damage [23]. To overcome these challenges, ablation methods utilizing alternative energy sources, such as cryoablation and PFA, have been developed to provide more effective and targeted treatment. Cryoablation procedures are associated with minimal endothelial disruption and negligible thrombus formation, enhancing their safety profile compared to traditional heat-based ablation techniques [24]. Additionally, reversible injury to cardiac tissue can be achieved when necessary, offering a unique advantage over non-reversible heat-based methods [25]. Nevertheless, reversible injury is a common limitation of cryoablation, often resulting in incomplete ablation and the need for repeat procedures to fully eliminate arrhythmogenic tissue [4]. To ensure effective treatment, sufficient cooling must be applied to achieve irreversible injury to the targeted tissue [25]. Additional risks of cryoablation include heart perforation, stroke, and damage to surrounding structures, such as the esophagus or pulmonary veins [25].

The success of RF ablation and cryoablation relies on accurate catheter placement, proper tissue contact, and controlled energy delivery. Moreover, using thermal energy can cause complications like cardiac tamponade, thromboembolism, pulmonary vein stenosis, and damage to surrounding structures [19]. To address these limitations, PFA has emerged as a potential alternative.

PFA is an innovative technique with high-voltage, short electrical pulses that induce irreversible electroporation in targeted tissue with minimal thermal damage while preserving surrounding structures [26]. Electroporation is a physical phenomenon that occurs when tissues are exposed to high-voltage electrical energy, creating temporary or permanent nanopores in cell membranes with minimal thermal effects [27]. PFA is characterized by superior thermal safety and efficacy compared to other ablation techniques that rely on thermal energy. Thermal energy in cryoablation and RF ablation can cause prolonged tissue damage due to residual heat persisting after the energy source is deactivated [27]. The ability to generate effective lesions in a tissue-selective manner may lead to meaningful improvements in the safety of procedures, reducing complications associated with thermal ablation such as atrial-esophageal fistula and hemi-diaphragmatic paralysis [28]. Moreover, the rapid nature of PFA can facilitate shorter ablation times and potentially improve patient throughput and reduce anesthesia exposure [29].

Despite PFA's advantages, this method faces several challenges that need to be addressed as technology advances. One concern is the lack of comprehensive long-term data on its safety and efficacy compared to established methods like RF ablation [30]. In addition, the absence of standardized equipment and protocols for PFA leads to variability in outcomes due to differences in energy settings and catheter designs [31].

The advantages, disadvantages, and complications of the three most common cardiac ablation methods are summarized in Table 1, highlighting the distinct benefits, risks, and effectiveness of RF ablation, cryoablation, and PFA across different patient profiles.

**Table 1. Clinical advantages and disadvantages of cardiac ablation methods**

| Cardiac ablation Method | Advantages | Disadvantages |
| --- | --- | --- |
| RF [123,124] | <ul><li>Lowest fluoroscopy time[*] [123]</li><li>Proven long-term efficacy in atrial fibrillation management [5]</li><li>Cost-effective compared to PFA or cryoablation [125]</li><li>Precise and controlled lesion formation [22]</li></ul> | <ul><li>Incidences (0.016%-0.1%) of esophageal injury [124]</li><li>Highest risk for pulmonary vein (PV) stenosis (incidence of severe PV stenosis: 0-0.5%) [124]</li></ul> |
| Cryoablation [7,25] | <ul><li>General anesthesia is not required due to the cooling effect [123]</li><li>Lower risk of thrombosis than RF [25]</li><li>Less patient discomfort [25]</li></ul> | <ul><li>Lowest acute success [123]</li><li>Higher recurrent arrhythmia rate than RF [25]</li><li>Incidences (0.08%-0.1%) of persistent phrenic nerve (PN) palsy following pulmonary vein isolation (PVI) [124]</li><li>Potential reversible damage [25]</li></ul> |
| PFA [19,26,27,29,126] | <ul><li>No esophageal complications, pulmonary vein stenosis, or persistent phrenic palsy [28]</li><li>Shortest procedure time [29]</li></ul> | <ul><li>Coronary arterial spasm and hemolysis-related acute renal failure [28]</li><li>Highest total costs [31]</li></ul> |

* Fluoroscopy involves real-time X-ray imaging during RF catheter ablation procedures [127].

The effectiveness of ablation techniques relies on balancing their benefits and limitations. Predicting potential complications before the procedure can significantly improve outcomes of cardiac ablation. Mathematical modeling is a tool that can be used to increase the success of cardiac ablation by improving preplanning, minimizing potential risks, and reducing failures, which ultimately lead to better patient outcomes.

### 3. Mathematical modeling

Mathematical modeling is essential for understanding and predicting outcomes and potential complications of cardiac ablations, such as heat-induced tissue changes and non-thermal effects, including tissue responses and interactions with electrical fields. Mathematical models are used to improve the safety and efficacy of cardiac ablation techniques by facilitating the design of catheters, optimizing the waveforms, and reducing the potential risks [15].

Pennes bioheat transfer equation, Maxwell's equations, and the Arrhenius equation are the main equations applied to model tissue damage and the complex interactions between thermal and electromagnetic phenomena within biological tissues during cardiac ablation [32]. The subsequent sections provide a detailed discussion of these equations and their applications in cardiac ablation.

### 3.1. Bioheat transfer models

Bioheat Transfer Models, such as Pennes bioheat equation [33], the dual-phase-lag model [34], and the Weinbaum-Jiji model [35] are developed for biological tissues, accounting for metabolic heat generation and perfusion effects. Bioheat Transfer models are commonly used in medical applications such as ablation [36]. Pennes bioheat equation is the primary model used to determine heat transfer within tissues, accounting for the effects of blood perfusion and metabolic heat generation (Eq. (1)) [37]:

$$\rho c_p \frac{\partial T}{\partial t} = \nabla(k\nabla T) + \rho_b \omega_b c_{pb}(T_a - T) + Q_m + Q_{ext} \tag{1}$$

Whereas $T$ (K) denotes the temperature of the tissue at a specified location and time, $k$ ($\frac{W}{m \cdot K}$) is the thermal conductivity of the tissue. $\rho$ ($\frac{kg}{m^3}$) and $c_p$ ($\frac{J}{kg \cdot K}$) represent the tissue density and specific heat capacity, respectively. $\omega_b$ ($\frac{1}{s}$) is the volumetric blood perfusion, accounts for metabolic heat generation within the tissue, while $Q_{ext} \left(\frac{\text{W}}{\text{m}^3}\right)$ represents the volumetric heat generation [38].

To derive the Pennes bioheat equation instantaneous thermal equilibrium between blood and tissue is assumed, which simplifies blood perfusion as uniform and local. Therefore, the dynamic and heterogeneous nature of vascular networks is neglected in the Pennes equation [39]. These limitations can lead to inaccuracies in predicting temperature distributions and thermal damage during ablation procedures.

Additionally, in the Pennes model, finite heat propagation speeds, thermal delays, non-Fourier heat transfer, and changes in tissue water content during overheating are overlooked. Finite heat propagation speeds and the interplay between blood flow and local tissue temperature are incorporated in the Weinbaum-Jiji and Dual-Phase-Lag models, offering a more accurate framework for predicting thermal energy distribution in biological tissues [34,35]. However, the inclusion of these terms significantly increases computational time while providing only marginal improvements in precision. Consequently, practical application of the Weinbaum-Jiji and Dual-Phase-Lag models remains limited. Despite addressing many of the Pennes bioheat equation's shortcomings, the trade-off between computational cost and accuracy makes the Pennes bioheat equation a more pragmatic and widely used choice for bioheat transfer modeling in ablation procedures [35].

Heating during RF ablation and PFA primarily results from the induction of an electromagnetic field within tissues, leading to Joule heating, potentially electrolysis, and plasma formation. Therefore, accurately modeling electromagnetic field distribution is crucial for precisely assessing tissue damage. The current electromagnetic models applied for cardiac ablation are discussed in the next section.

### 3.2. Electromagnetic models

Thermal modeling improves the understanding of the heat-related effects in PFA, RF ablation, and cryoablation. RF ablation and PFA techniques use electrical energy to trigger cellular effects; therefore, electromagnetic simulations are required to accurately model the interaction between electromagnetic fields and biological tissues [40]. RF ablation uses high-frequency alternating currents to cause thermal damage [41], whereas PFA employs short, high-voltage pulses to disrupt cell membranes with minimal heat [42]. Developing electromagnetic models is essential for predicting lesion size, clarifying ablation mechanisms, and optimizing procedural parameters, ultimately improving clinical outcomes and safety [43].

Electromagnetic-thermal coupled models integrate Maxwell's equations for electromagnetic fields with bioheat transfer equations to predict the spatial and temporal distribution of temperature and electric potential in the cardiac tissue surrounding the ablation zone [40]. Maxwell's equations are coupled with the Pennes bioheat equation and the Navier-Stokes equations to model the complex interactions between electromagnetic fields, heat transfer, and fluid dynamics, forming a comprehensive Multiphysics simulation [44]. Moreover, the specific absorption rate (SAR) is utilized in these simulations to quantify the heating effects generated by electromagnetic fields, ensuring accurate predictions of thermal responses and lesion formation [45]. The electromagnetic-thermal coupled models are used to determine the impact of different ablation parameters, such as discharge time [40], discharge voltage [46], and electrode size [47] on the temperature distribution. The electric and magnetic fields distribution based on the input voltage and device geometry are included in Maxwell's equations (Eqs. (2) to (4)) [48].

$$\nabla \cdot (\sigma \nabla V) = 0 \tag{2}$$

$$E = -\nabla V \tag{3}$$

$\sigma \left(\frac{S}{m}\right)$ is the electrical conductivity of the tissue, $V$ (V) is the electric potential, and $E \left(\frac{V}{m}\right)$ is the electric field strength [48]. The heat source term is then given by the Joule heating equation (Eq. (4)) [49].

$$Q = \sigma |E^2| \tag{4}$$

Where $Q \left(\frac{W}{m^3}\right)$ is the volumetric generated heat. This equation is crucial for simulating the heating patterns in RF ablation and PFA [50].

The fundamental principle of RF ablation involves generating resistive heating in cardiac tissues through the application of alternating current with frequencies between 300 kHz and 1 MHz [51]. As the current passes through the tissue, it causes water molecules near the electrode to vibrate, resulting in energy deposition, leading to thermal effects and subsequent cellular death [52]. Tissue near the electrode is heated due to the Joule heating effect, while the temperature in more distant areas primarily rises through thermal conduction and convection [53]. The volume of the lesion can be determined by considering the area that reaches temperatures above 50 °C [54]. RF ablation is governed by the Laplace equation for electric field distribution and the bioheat equation for modeling temperature changes during the procedure (Eq. (5)) [52]:

$$\nabla[\sigma(T)\nabla V] = 0 \tag{5}$$

$\sigma(T) \left(\frac{S}{m}\right)$ is the temperature-dependent electric conductivity, and $V$ (V) is the electric potential.

Eq. (5) is incorporated a temperature-dependent conductivity term multiplied by the electric potential squared, which allows for a more accurate representation of energy absorption and temperature increase as tissue conductivity varies with temperature [55]. The temperature can be determined by solving the Pennes bioheat equation, as shown in Eq. (1) [52]. This approach enhances predictive accuracy regarding lesion formation and depth; however, it introduces computational complexity and requires precise data on tissue properties [55].

Joule heating can be included by multiplying current density by electric field intensity, offering a simpler and faster computational framework suitable for real-time applications (Eq. (6)) [56]:

$$\rho c_p \frac{\partial T}{\partial t} = \nabla(k \nabla T) + J * E - Q_h \tag{6}$$

Where $J\left(\frac{A}{m^2}\right)$ is the current density, $E\left(\frac{V}{m}\right)$ is the strength of the electric field, and $Q_h\left(\frac{W}{m^3}\right)$ accounts for volumetric heat loss due to blood perfusion in the myocardial wall, which can be neglected since it is very small in comparison with other terms in Eq. (6) [57]. Crucial thermal interactions such as variable blood perfusion and velocity, two-phase water vaporization, local thermal non-equilibrium between tissue and blood phases, changes in thermal conductivity, and anisotropy of thermal properties are neglected in this model, which limits the accuracy of the thermal damage and lesion characteristics calculated using Eq. (6) [58].

PFA is distinct from traditional techniques like RF ablation, relying on high-voltage pulsed electric fields instead of alternating current [58]. Eqs. (1) and (2) are fundamental in simulating the electrical and thermal effects of PFA by modeling the electric field distribution and the energy deposition in biological tissues. Most current numerical models for predicting lesion formation during PFA are based on the principle that lesions occur when the electric field intensity exceeds a specific threshold, known as the irreversible electroporation threshold [59]. This threshold can be affected by various factors, including the distance of the catheter from the tissue [60], pulse duration, waveform characteristics, the number of pulses delivered, and the intervals between successive pulses [61]. A quasi-static model with steady-state electric field simulation and time-dependent thermal analysis can reduce PFA modeling computational costs [62]. In this framework, it is assumed that the current density within the tissue is divergence-free, meaning there is no net accumulation or depletion of electric charge during each pulse. Mathematically, this condition is expressed as $\nabla \cdot J = 0$, where $J\left(\frac{A}{m^2}\right)$ represents the current density vector. This assumption is often used in quasi-static conditions to model electric field distribution and tissue interaction during electroporation [45]. Changes in tissue electrical properties, such as conductivity and permeability, affect the electric field distribution within the system. Consequently, assuming a divergence-free field introduces inaccuracies in predicting thermal damage and electroporation-induced cell death. The mathematical models used to calculate lesion size resulting from thermal and electroporation damage are presented in Section 3.3.

### 3.3. Thermal and electroporation-induced tissue damage models

Arrhenius equation is widely used to predict tissue thermal damage based on the cumulative effect of temperature and heating duration [63]. The Arrhenius thermal damage equation is used to calculate the probability of cell death by integrating the exposure of tissue to elevated temperatures over time [64]. A key advantage of the Arrhenius approach is its ability to model tissue damage at various temperatures, making it a flexible and powerful method in medical procedures [65]. Arrhenius's equation is expressed as follows (Eq. (7)) [64]:

$$\Omega = A exp\left(-\frac{E_a}{RT}\right)\Delta t \qquad (7)$$

Where $\Omega$ is the thermal damage function, $A\left(\frac{1}{s}\right)$ is the pre-exponential factor constant, a tissue-specific parameter that varies based on experimental conditions, $E_a\left(\frac{J}{mol}\right)$ is the activation energy, $R\left(\frac{J}{mol}\right)$ is the gas constant, $T$ (K) is the temperature, and $\Delta t$ (s) is the time increment. Eq. (8) is used to derive a percentage value representing cell death from thermal damage [66]:

$$ThermalDamage_{kill} = 100 * (1 - \exp(-\Omega(t))) \qquad (8)$$

Another model for predicting hyperthermic injury is the Thermal Iso-effective Dose (TID) [67] or cumulative equivalent minutes at 43 (CEM$_{43}$) [67–72]. The thermal equivalent minutes approach is used to determine how long a specific tissue can be maintained at a given temperature before damage occurs. Most types of tissue generally begin to experience damage at 43°C, making this temperature a critical reference point [73]. This model is commonly used to identify the heating duration required to cause thermal tissue damage and is often utilized to set safe exposure thresholds [74]. It allows non-isothermal heating conditions to be compared to isothermal heating at a reference temperature, typically set at 43°C (Eq. (9)) [69].

$$TID \text{ (or } CEM_{43}) = \int_0^t C^{(43-T(t))}dt, \quad \begin{cases} C = 0.25, & T < 43 \cdot C \\ C = 0.5, & T \geq 43 \cdot C \end{cases} \qquad (9)$$

In this model, $T(t)$ indicates the temperature applied to the target tissue at each instant, $dt$ represents the time spent (min) at a certain temperature $T$ (°C), and $C$ is an adjustment factor for each 1°C change in temperature [75]. This parameter ($C$) is commonly represented as $R$ in the relevant literature. However, since we used $R$ as the gas constant in Eq. (7), we chose an alternative notation to avoid confusion [76]. In most soft tissues, the coagulative necrosis threshold ranges between 100 and 1000 minutes at 43°C [77].

Thermal exposure and electric field distribution are key factors in lesion formation during ablation. When the transmembrane potential surpasses a critical threshold (750 $\left(\frac{V}{cm}\right)$ [76] - 1000 $\left(\frac{V}{cm}\right)$ [61] for 100μs pulses), depending on the waveform configuration, irreversible electroporation causes permanent membrane disruption and cell death [78]. Understanding these dynamics is essential for analyzing lesions created by PFA ablation, as the applied electric field directly impacts the extent of cell destruction. The ratio of surviving cells after electroporation to the number of cells before treatment can be determined by Eqs. (10) to (13) [66,79]:

$$S = \frac{1}{1 + \exp\left(\frac{E - E_c(n)}{B(n)}\right)} \quad (10)$$

$$E_c(n) = E_{c0} \exp(-g_1 n) \quad (11)$$

$$B(n) = B_0 \exp(-g_2 n) \quad (12)$$

$$Electroporation_{kill} = 100 * (1 - S) \quad (13)$$

Where, $S$ is the ratio of surviving cells after electroporation, $E$ $\left(\frac{V}{m}\right)$ and $E_c(n)$ $\left(\frac{V}{m}\right)$ denote the applied electric field and the critical electric field, respectively, which $E_c(n)$ constitutes the critical electric field necessary for the death of 50% of the cell population. $B(n)$ $\left(\frac{V}{m}\right)$ is a variable that depends on the number of pulses delivered [66]. $E_0$ $\left(\frac{V}{m}\right)$, $B_0$ $\left(\frac{V}{m}\right)$, $g_1$, and $g_2$ are constant values determined through regression analysis. The regression analysis conducted utilized electroporation properties, highlighting the scarcity of experimental data for specific tissues, including cardiac tissue [66,79,80].

Electroporation-induced and thermal cell damage, along with lesion size, are evaluated using Eqs. (7) to (13). Applying these equations does not account for the synergistic effects of thermal and electroporation-induced damage, which may lead to inaccuracies in predicting the actual survival rate [66]. In addition, Eq. (13) is derived empirically for electric pulses with microsecond durations and calibrated using data from liver tissue rather than myocardial tissue. As a result, the constants and coefficients are only valid for pulses within the microsecond range and cannot be applied to shorter or longer pulse durations. Therefore, recalibration of Eq. (13) for myocardial tissue is necessary. Other models relied on their own data set for calibration, specific to their PFA waveform and catheter geometry [61].

## 4. Electrical and thermal properties of cardiac tissue

Heat transfer in cardiac tissues depends on density, thermal conductivity, and specific heat capacity [81]. Myocardial tissue generally has higher thermal conductivity compared to epicardial tissue, which influences the distribution of heat during ablation procedures [81]. The effective thermal conductivity of the myocardium decreases when myocardial tissue is surrounded by fibrous and adipose tissues, and as a result, it slows down the heat transfer rate from the myocardium to the next layer of cardiac tissues [82]. Although certain models consider the dependence of properties on myocardial fiber orientation [83], the heterogeneous nature of cardiac tissue is frequently overlooked due to its inherent complexity [84].

Thermal and electrical tissue properties are determined through experimental techniques. When direct measurement is impractical, simulations and mathematical modeling serve as complementary tools to predict the impact of these properties on final outcomes of cardiac ablation. Ex vivo measurement of thermal properties is often unreliable due to lack of perfusion [81]. Statistical information found in databases can provide valuable insights into the variability of tissue properties [85]. A list of the thermal and electrical property values incorporated in cardiac ablation computational models is presented in Table 2.

It should be noted that the properties of cardiac tissue are affected by temperature, pressure, electrical field, and the type of tissue involved [84]. For instance, an increase in temperature during cardiac ablation enhances tissue thermal conductivity, and the electrical field can alter the electrical conductivity of the tissue [86].

**Table 2. Electrical and thermophysical properties of cardiac tissues and ablation catheter**

| Material | $\sigma_0\ (\frac{S}{m})$ | $\sigma_1\ (\frac{S}{m})$ | $k\ (\frac{W}{m\cdot K})$ | $\rho\ (\frac{kg}{m^3})$ | $c_p\ (\frac{J}{kg\cdot K})$ |
|---|---|---|---|---|---|
| Electrode | 4.6 x10$^6$ [86] | - | 71 [86] | 21,500 [86] | 132 [86] |
| Catheter | 10$^{-5}$ [86] | - | 23 [86] | 1440 [86] | 1050 [86] |
| Saline water | 1.392 [86] | - | 0.628 [86] | 980 [86] | 4184 [86] |
| Epicardial fat/ adipose tissue | 0.0377 [86]<br>0.0684 [85] | 0.0438 [86] | 0.21 [85] | 911 [85] | 2348 [85] |
| Heart/ myocardium | 0.0537 [86]<br>0.733 [85] | 0.281 [86] | 0.56 [85] | 1081 [85] | 3686 [85] |
| Cardiac chamber/ blood | 0.7 [86]<br>1.23 [85] | 0.748 [86] | 0.52 [85] | 1050 [85] | 3617 [85] |
| Connective tissue* | 0.1199 [86]<br>0.490 [85] | - | 0.35 [86]<br>0.39 [85] | 1000.5 [86]<br>1027 [85] | 2884.5 [86]<br>2372 [85] |

$\sigma$: electrical conductivity ($\sigma_0$ and $\sigma_1$ are the pre- and post-electroporation conductivity values, respectively); $k$: thermal conductivity; $\rho$: density; $c_p$: specific heat
*Mixture of 50% fat and 50% muscle

Once the mathematical model and thermal-electrical properties are defined, the next step in simulating the cardiac ablation procedure is selecting an appropriate computational method to solve the governing equations and predict treatment outcomes, as discussed in the following sections.

## 5. Computational modeling approaches

Computational modeling is essential for predicting tissue damage during ablation procedures, which helps to predict treatment outcomes and optimize treatment parameters [87]. The choice of computational method for ablation modeling depends on the complexity of the procedure, the required precision, and the available computational resources [88]. The most common computational approaches in ablation involve applying numerical methods to solve momentum, mass, and energy balance equations [89] and/or utilizing Artificial Intelligence (AI) to develop mathematical correlation models trained on existing data (Table 3) [90,91].

The interaction between tissue, ablation catheter, and blood flow during ablation may significantly impact heat dissipation and lesion formation, potentially altering the depth and size of the lesions [92]. During atrial fibrillation ablation procedures, computational fluid dynamics (CFD) method is particularly valuable for modeling blood flow within the heart's chambers and evaluating the cooling effects of blood flow to ensure effective lesion formation [93].

In addition, development of patient-specific models is facilitated by recent advancements in computational methods, enabling individualized treatment planning for both atrial and ventricular ablation procedures [94].

Several numerical methods are used in ablation modeling, each suited to specific scenarios. Finite Element Method (FEM) is widely used because it can provide detailed simulations, though it can be computationally expensive for large-scale or transient simulations [95]. FEM is utilized to simulate thermal and mechanical responses during ablation and is commonly applied for its flexibility in handling complex geometries and boundary conditions, especially in RF ablation [96].

The Finite Difference Method (FDM) is simpler to implement and requires fewer computational resources, but it has limitations with complex geometries and provides less accurate results. FDM can be applied effectively for one- or two-dimensional problems, such as thermal diffusion along a linear catheter path [97].

The Finite Volume Method (FVM) is another numerical technique for modeling ablation problems, particularly when blood flow plays a crucial role in thermal ablation scenarios. FVM is used to simulate various aspects of ablation, including the blood velocity fields and friction coefficient variations by ensuring the conservation of mass, momentum, and energy within the control volumes [98]. FVM has been applied to various ablation applications, including RF ablation and cryoablation, enabling the simulation of ablation zones, temperature distributions, and ablation efficiency [99].

The Lattice Boltzmann Method (LBM) is employed to solve the governing equations involved in complex heat-fluid interactions, such as tissue vaporization during ablation. It provides a numerical framework for fluid dynamics at the macroscopic scale, based on kinetic equations formulated at the mesoscopic scale [100]. Mono-domain cardiac electrophysiology can be efficiently simulated using LBM [101]. LBM's capacity in handling complex geometries and boundary conditions makes it ideal for modeling the detailed aspects of PFA procedures [102].

In the context of ablation, AI techniques, especially Machine Learning (ML), can be applied to optimize treatment protocols by extracting significant parameters that impact the treatment outcomes from the existing patient data [103]. AI-based models are increasingly being incorporated into ablation modeling to optimize parameters and predict real-time outcomes. Algorithms in the ML methods are used to optimize ablation strategies (e.g., placement of probes) to achieve desired outcomes with minimal damage to healthy tissue [104]. For instance, clinicians can use ML models to process intraoperative data for monitoring and predicting ablation zone growth. The macro-classification method in ML has been applied to RF ablation [105]. ML models have been used to predict outcomes using clinical data and electrograms, offering insights into tissue damage during the freezing process [106]. Integrating electrogram and electrocardiogram signals with clinical data through ML can improve predictions of atrial fibrillation recurrence after PFA ablation [107].

In summary, FEM is the primary numerical method for accurately simulating thermal damage in RF and cryoablation, while FDM is used for fast response and simplified 1D and 2D computational models. ML methods are growing in use for predicting patient-specific outcomes and enabling personalized treatments. FVM and LBM are utilized in atrial and ventricular applications where blood flow significantly impacts ablation outcomes (Table 3). Despite advances in computational models for predicting cardiac ablation outcomes, the development and validation of multiphysics, patient-specific simulations capable of accurately predicting lesion size and tissue damage remain active areas of research, particularly for emerging techniques like PFA. A major challenge in this field is designing a computational algorithm that can simultaneously solve the Navier-Stokes, Maxwell, and Pennes equations in complex geometries. Developing such algorithms is time-intensive and requires highly skilled experts. To expedite this process, commercially available software provides user-friendly interfaces and preprogrammed solvers, reducing the effort needed for algorithm development. The next section discusses the available software options for these simulations.

**Table 3. Computational Modeling for atrial and ventricular ablation**

| Geometry | Computational modeling method | Ablation method | Ablation Pattern | Is thermal damage assessed? | Is electroporation assessed? |
|---|---|---|---|---|---|
| Atrial | FEM | RF | Phase singularity-based and Dominant Frequency (DF)-based [128] | × | × |
| | | | Focused on rotor ablation using basket catheter strategies [129] | × | × |
| | | | Focus on lesion modeling [130] | √ | × |
| | | | Pulmonary Vein Isolation (PVI) [131] | × | √ |
| | | PFA | Dose-dependent lesion depth correlated with voltage and tissue contact [60] | × | √ |
| | | | Focus on lesion modeling (thermal and IRE ablation effects) [89] | √ | √ |
| | | | PVI [132] | × | √ |
| | | Cryoablation | PVI [133] | √ | × |
| | | | PVI [134] | √ | × |
| | FDM | RF | PVI [135] | × | × |
| | | | DF-based ablation [136] | × | × |
| | | | Reentrant driver defined by 3D structural "fingerprints" in Atrial Fibrillation (AF) [137] | × | × |
| | ML | RF & Cryoablation | PVI and additional ablation lines based on clinical need [138] | × | × |
| | | | PVI [139] | × | × |
| | | | PVI [140] | × | × |
| Ventricular | FEM | RF | No specific ablation pattern mentioned [141] | √ | × |
| | | | Substrate-based ablation for ventricular tachycardia [91] | × | × |
| | | | Focuses on computational lesion modeling [121] | √ | × |
| | | PFA | Focus on lesion modeling (optimization of IRE protocols for myocardial decellularization and damage control) [142] | √ | √ |
| | | PFA & RF | No specific ablation pattern mentioned, focused on lesion morphology [61] | √ | √ |
| | ML | RF | Substrate-based ablation [143] | × | × |

×No    √ Yes

## 6. Computational modeling platforms for cardiac ablation simulations

COMSOL Multiphysics is widely used for modeling ablation due to its multiphysics capabilities, enabling the integration of electromagnetic, bioheat transfer, and fluid dynamics modules [108]. Researchers use COMSOL Multiphysics to incorporate temperature-dependent properties, model multiple physics interactions simultaneously, and integrate user-defined functions [108]. COMSOL is utilized to model temperature distribution within the ablation catheter and cardiac tissues during PFA and to evaluate the effects of pulse number and electrical conductivity on cell ablation and thermal damage [66]. However, integrating multiple physics, such as electromagnetic, bioheat transfer, and fluid dynamics, significantly increases the complexity of setting up and running simulations in COMSOL, primarily due to its FEM-based approach. Thus, when blood flow plays a critical role in determining lesion size, COMSOL may not be the most suitable option.

ANSYS provides extensive tools for fluid dynamics, thermal, and electromagnetic simulation, making it a desirable tool for modeling PFA [109]. ANSYS electromagnetic module, high-frequency structure simulator (HFSS), is advantageous for high-frequency applications like RF ablation. Additionally, ANSYS transient thermal module can be used to incorporate temperature-dependent blood perfusion in thermal modeling [110]. Temperature-dependent properties can be incorporated into ANSYS models to enhance the precision of electrode design optimization and ablation parameter determination [111]. The learning curve for ANSYS is steeper than COMSOL, as COMSOL offers a highly intuitive and user-friendly interface. Additionally, performing thermal analysis and post-processing in ANSYS is more complex, requiring a highly experienced user. In complex cases, programming languages and numerical computing environments such as MATLAB and Python are used to develop customized models, either independently or in conjunction with software like COMSOL and ANSYS.

MATLAB is utilized for modeling and simulating cardiac ablation, enabling temperature distribution control, energy delivery optimization, and lesion characterization under varying blood flow conditions [46,63]. It is also used to analyze catheter position and stability, supporting improved lesion quality and procedural outcomes [112]. Both MATLAB and COMSOL have shown high applicability in assessing tissue responses to cardiac ablation, with some studies integrating both platforms for a comprehensive approach to refining ablation strategies [113].

Python tools are utilized for creating sophisticated simulations and ML models for cardiac ablation application [114,115]. Applications such as predicting cardiac ablation outcomes and optimizing procedural strategies demonstrate Python's versatility in analyzing complex medical data and training advanced models [116].

Cardiac ablation simulations may not reflect real-life outcomes without validation and verification. Therefore, systematic verification and validation are essential to ensure the reliability of computational models. The following section discusses the validation process and recent research addressing this need.

## 7. Validation of cardiac ablation simulations

Validation is primarily used to refine and improve a computational model to ensure it accurately represents reality. Once this is achieved, the validated model can be trusted for use in the decision-making process and clinical settings, providing reliable insights to guide treatments and procedural strategies [117]. A key focus of recent research is improving the precision and effectiveness of cardiac ablation simulations by comparing their results with clinical outcomes in cardiac ablation [118].

Only a limited number of cardiac ablation models have been validated using experimental data. For instance, a model of irrigated RF ablation was validated by comparing temperature, lesion width, and depth between simulations and experiments [119]. In this study, for perpendicular catheter orientation, errors were 6.2°C for maximum gel temperature, 0.7 mm (10.9%) for lesion width, and 0.3 mm (7.7%) for lesion depth [119]. These errors highlight the limitations of model accuracy in predicting thermal effects during cardiac ablation. While the discrepancies prevent full confidence in the computational model, they help identify inaccuracies and guide model improvements.

Open-irrigated electrodes optimize power delivery while maintaining low temperatures. As another example, the computational model of ThermoCool (6-hole) and ThermoCool® SF (multi-hole) catheter electrodes for endocardial RF ablation were validated against experimental lesion dimensions [120]. Differences between computational model and experiment in lesion depth were below 1 mm, while lesion width varied within 1-2 mm, following 30 and 60 seconds of RF ablation at 20W and 35W [120]. Based on the validation results, the model accurately predicts lesion depth within acceptable limits, while deviations in lesion width highlight the need for further refinement to improve accuracy [120].

Lesion size depends on the power dissipated in the tissue, which is influenced by the electrode's contact area [121]. In a computational model of radiofrequency catheter ablation with open-irrigated electrodes, lesion depth was validated by comparing computational model results with *in vitro* porcine myocardium experiments, yielding errors from -1.16% to +5.42% [121]. The model underestimated lesion width by 9–23% and predicted the maximum lesion width at a greater depth than observed experimentally, overestimating it by up to 52%. This suggests the computational model may overestimate heat penetration into deeper tissue, possibly due to assumptions regarding thermal conductivity, tissue perfusion, or energy distribution [121]. These results highlight the need for further model improvements, such as incorporating tissue structure, to enhance accuracy [121].

These examples underscore the criticality of validating and refining computational models to achieve reasonable accuracy before their application in clinical decision-making. Validation of cardiac ablation models with *in vivo* experimental data is challenging due to tissue variability, the heart's dynamic environment, and difficulties in real-time measurement. Indirect imaging methods have limitations, and ethical and logistical constraints make *in vivo* testing costly and complex [122]. *In vitro* models serve as alternative systems for collecting experimental data but lack the effects of perfusion, cardiac dynamics, and thermophysical property variations. Consequently, many studies rely on computational modeling without comprehensive validation and verification. Validation and verification of computational models are critical across engineering disciplines, with standards such as ASME V&V 20-2009 and ASME V&V 40-2018 providing guidelines for biomedical applications.

In summary, a key knowledge gap in computational cardiac ablation is the lack of experimental data for validating simulation results. Potential solutions include developing non-invasive measurement techniques, utilizing AI to extract data from current measurement modalities such as MRI, CT, ultrasound, and echocardiography, and fabricating realistic *in vitro* models using advanced organoid and tissue printing technologies.

## 8. Conclusions and future directions

A comprehensive review of computational methods used in cardiac ablation treatments is provided in this paper, offering new insights to propel future research in the field. Physics-driven models alongside emerging AI-based models are discussed in this paper. The significant potential of computational modeling to enhance the planning, execution, and prediction of outcomes in cardiac ablation procedures is highlighted in this paper.

Despite the advancements in cardiac ablation modeling, a significant challenge remains the scarcity of comprehensive *in vivo* experimental data for thorough validation. The limited availability of detailed lesion measurements in various cardiac tissues and ablation techniques hinders the refinement and validation of these models. To address this, future research should prioritize the acquisition of high-quality experimental data, including detailed lesion size, depth, and transmural measurements, across a range of ablation parameters and tissue types.

To transition computational models from experimental research to routine clinical practice, it is essential to integrate patient-specific models with lesion prediction models to enhance accuracy and clinical applicability. Patient-specific models are constructed using high-resolution intraoperative imaging and post-procedural histopathology. By incorporating patient-specific anatomical and physiological data into lesion prediction simulations, clinicians can dynamically refine ablation strategies to optimize lesion placement and minimize collateral damage.

Integrating machine learning with computational fluid-thermal-electrical models enables real-time simulation of cardiac ablation, supporting clinical decision-making and adaptive treatment strategies based on patient-specific responses. Transitioning these models from research to routine clinical practice requires large-scale, multi-center validation studies to ensure predictive accuracy and safety.

**FUNDING**

The authors gratefully acknowledge the support of Fields Medical, Inc. [Grant No: GFP00144], and the University of North Texas for funding this research.

**Conflict of interest:** none declared.

**DATA AVAILABILITY**

No new data were generated or analyzed in support of this research.

**Author contributions**

**Leila Seidabadi** and **Indra Vandenbussche:** Contributed equally to the conceptualization; Literature review; Writing-original draft; Supervision; Writing-review and editing. **Rowan Carter Fink**: Contributed to data collection, visualization; writing-original draft. **MacKenzie Moore**: Contributed to literature review; writing-review and editing. **Bailey McCorkendale:** Contributed to literature review; writing-review and editing. **Fateme Esmailie**: Contributed to supervision; conceptualization; resources; writing-review and editing. All authors read and approved the final manuscript.